\documentclass[final]{aipproc}
\layoutstyle{6x9}
\usepackage{natbib}
\usepackage{amsmath}
\usepackage{amsfonts}
\usepackage{amssymb}
\usepackage{multirow}
\usepackage{color}
\usepackage{colortbl}

\def\apj{Astrophys. J.}
\def\apjl{Astrophys. J. Lett.}

\def\aap{Astron. Astrophys. }

\def\physrep{Phys. Rep. }
\def\mnras{Mon. Not. Roy. Astron. Soc. }

\def\prl{Phys. Rev. Lett.}
\def\prd{Phys. Rev. D.}

\def\cqg{Class. Quant. Grav.}

\begin{document}

\title{Studies of Stellar Collapse and Black Hole Formation with the Open-Source Code GR1D}

\classification{04.25.D-, 04.40.Dg, 97.10.Kc, 97.60.Bw, 97.60.Jd, 97.60.Lf, 26.60.Kp}
\keywords      {stellar collapse, neutron stars, black holes, supernovae}

\author{C. D. Ott}{
  address={cott@tapir.caltech.edu, TAPIR, California Institute of Technology,
  Pasadena, CA, 91125, USA}
}

\author{E. O'Connor}{
  address={evanoc@tapir.caltech.edu, TAPIR, California Institute of Technology,
  Pasadena, CA, 91125, USA}
}

\begin{abstract}
We discuss results from simulations of black hole formation in failing
core-collapse supernovae performed with the code GR1D, a new
open-source Eulerian spherically-symmetric general-relativistic
hydrodynamics code. GR1D includes rotation in an approximate way
(1.5D),  comes with multiple
finite-temperature nuclear equations of state (EOS), and treats
neutrinos in the post-core-bounce phase via a 3-flavor leakage scheme
and a heating prescription. We chose the favored $K_0 =
220\,\mathrm{MeV}$-variant of the Lattimer \& Swesty (1990) EOS and
present collapse calculations using the progenitor models of Limongi
\& Chieffi (2006). We  show that there is no direct (or ``prompt'') black
hole formation in the collapse of ordinary massive stars ($8\,M_\odot
\lesssim M_\mathrm{ZAMS} \lesssim 100\,M_\odot$) and present
first results from black hole formation simulations that include rotation.
\end{abstract}

\maketitle


\section{Introduction}

Core-collapse supernova explosions are the dramatic events heralding
massive star death in core collapse.  All stars in the
zero-age-main-sequence (ZAMS) mass range from $\gtrsim 8\,M_\odot$ to $\sim
100\,M_\odot$ undergo core collapse at the end of their life, but not
all core collapse events result in a core-collapse supernova
explosion. In any star of the mass range under consideration, core
collapse separates the stellar core into subsonically collapsing inner
core and supersonic outer core. The collapse of the former is
stabilized by the stiffening of the nuclear equation of state (EOS)
near nuclear density.  Core bounce occurs, launching a hydrodynamic
shock into the still infalling outer core. This prompt shock fails to
blow up the star and is forced into stall by the dissociation of
accreting iron-group nuclei and neutrino losses from the postshock
region. The shock must be \emph{revived} for core collapse to result
in an explosion. The precise mode of revival provided by the much
sought-after supernova mechanism is still uncertain
(e.g., \cite{janka:07,marek:09,ott:08,burrows:06,
  burrows:07b,sagert:09}), and in a finite (though unknown) fraction
of massive stars, it must fail completely to drive an explosion, or,
at least fail to unbind the entire star so that significant fallback
accretion occurs \cite{timmes:96}. In the former case, a black hole
(BH) inevitably forms within a few seconds as the stellar mantle
accretes onto the protoneutron star (PNS; Collapsar type
I~\cite{heger:03}), while in the latter, a neutron star initially
survives but may be pushed over its mass limit by fallback accretion
(Collapsar type~II~\cite{heger:03,zhang:08}). Both are considered as
potential scenarios leading to a long gamma-ray
burst (GRB)~\cite{heger:03,wb:06}.

BH formation in the core collapse context has been studied with full
general-relativistic (GR) Boltzmann neutrino radiation-hydrodynamics
codes in spherical symmetry~without rotation (e.g.,
\cite{fischer:09a,sumiyoshi:06,sumiyoshi:07} and references therein),
but due to the computational cost of these high-fidelity simulations,
only few models have been simulated. Multi-dimensional simulations of
BH formation remain to be performed with appropriate microphysics and
progenitor models, but see \cite{sekiguchi:05} for an exploratory work
with simplified physics.

We have recently implemented the new code GR1D, an Eulerian
spherically-symmetric code for stellar collapse and BH
formation~\cite{oconnor:10}. GR1D is \emph{open-source} and may be
downloaded from {\tt http://www.stellarcollapse.org}. GR1D, in its
present version, does not implement the full radiation transport
formalism, but instead relies on an approximate, yet extremely
computationally efficient leakage scheme for neutrino transport~(e.g.,
\cite{ruffert:96,rosswog:03b})
and a simple prescription for
neutrino heating. Moreover, GR1D
implements an approximate way of including rotation in spherical
symmetry a variant of which is used in stellar evolutionary
calculations~(e.g., \cite{heger:00}).
GR1D's computational efficiency allows us to perform 
hundreds of model simulations in little time to explore
systematically the conditions for BH formation in massive stars and
the parameter space in ZAMS mass, metallicity, and rotation where BH
formation may be the dominant outcome of stellar collapse. Detailed
results of such an extensive study will be reported in \cite{oconnor:10b}.

In the following, we discuss some details of the GR1D code, then
present simulation results highlighting the fact that any BH
forming core collapse passes through a PNS phase and is never
``direct''. We then show first results from simulations of
rotating BH formation.

\section{The GR1D Code}
GR1D follows the $3+1$ approach to numerical relativity, slicing 4D
spacetime into 3D spacelike hypersurfaces along a timelike normal
(e.g., \cite{alcubierre:08}). This introduces two gauge quantities,
the lapse function $\alpha$ and the shift vector $\vec{\beta}$ whose
choices are not a priori fixed. The lapse controls how time changes
between two consecutive slices while the shift describes how
coordinates change from one slice to the next. In GR1D, we adopt
spherical symmetry and the polar-slicing, radial-gauge choice~(e.g.,
\cite{gourgoulhon:91}) on an Eulerian grid, resulting in $\vec{\beta}
= 0$ and a Schwarzschild-like invariant line element of the form
(assuming $c=G=M_\odot=1$ here and in the following),
\begin{equation}
ds^2  =  -\alpha(r,t)^2 dt^2 + X(r,t)^2 dr^2 + r^2 d\Omega^2 \,\,,
\end{equation}
where the lapse $\alpha$ and the metric function $X$ are functions of
a metric potential $\Phi(r,t)$ and of the enclosed gravitational
mass $M_\mathrm{grav}(r,t) = m(r,t)$,
\begin{equation}
\alpha(r,t) = \exp\left[\Phi(r,t)\right],\hspace*{1cm}
X(r,t) = \left( 1 - {2 m(r,t) \over
    r}\right)^{-1/2}\,\,.
\label{eq:metriccoefficients}
\end{equation}
The detailed form of $\Phi(r,t)$ can be found in
\cite{oconnor:10,gourgoulhon:91}. Here we point out only that the
choice of $\Phi$ ensures that $\alpha$ is singularity-avoiding and,
hence, drops to very small values when a physical singularity forms, thus
minimizing its evolution with coordinate time. It is important to note
that the metric function $X$ becomes singular at $r=2M$ and thus
forbids numerical evolution beyond BH formation in this gauge (a
feature in common with many other 1D
codes~\cite{fischer:09a,sumiyoshi:07}).

GR1D's hydrodynamics module follows the flux-conservative Valencia
formulation of GR hydrodynamics in the form of~\cite{romero:96}. The
scheme is semi-discrete in space and is finite-volume with
piecewise-parabolic reconstruction of interface values and employs the
HLLE Riemann solver \cite{HLLE:88}. Time discretization is handled via
2nd- or 3rd-order Runge-Kutta integrators using the Method of
Lines~\cite{Hyman-1976-Courant-MOL-report}.

A specialty of GR1D is its approximate inclusion of rotation in
spherical symmetry (1.5D). This is accomplished by solving an
advection equation for the angular momentum and including an angularly
averaged centrifugal term in the radial momentum equation, in the
Lorentz factor, and in curvature terms to account for centrifugal
force, angular momentum flux, and rotational energy~\cite{oconnor:10}.

GR1D operates with a general EOS interface and has been tested with
variants of the Lattimer-Swesty EOS~\cite{lseos:91} and the
H.~Shen~EOS~\cite{shen:98b} tables of which we make available
for download at {\tt http://stellarcollapse.org} and describe in \cite{oconnor:10}.

Deleptonization and neutrino transport are handled by GR1D differently
in the prebounce and postbounce phases. In the former, we employ the
simple parameterization of the electron fraction
$Y_e$ as a function of rest-mass density $\rho$ put forth by
\cite{liebendoerfer:05fakenu}. In the latter, we use a 3-flavor ($\nu_e$,
$\bar{\nu}_e$, and $\nu_x = \{\nu_\mu,\bar{\nu}_\mu,
\nu_\tau,\bar{\nu}_\tau\}$) energy-averaged leakage scheme constructed
along the lines of \cite{ruffert:96,rosswog:03b} and described in
\cite{oconnor:10}. It yields energy-integrated neutrino luminosities
that are within $\sim 20\%$ of those predicted by full Boltzmann
transport calculations. Neutrino heating by charged-current absorption of
$\nu_e$ and $\bar{\nu}_e$ is
handled via 
\begin{equation}
Q^{\mathrm{heat}}_{\nu_i}(r) = f_\mathrm{heat} \frac{L_{\nu_i}(r)}{4\pi r^2}
\sigma_{\mathrm{heat},\nu_i}\, {\rho\over m_u} X_i \left\langle {1 \over
  F_{\nu_i}} \right\rangle e^{-2\tau_{\nu_i}} \,\,,
\end{equation}
where $L_{\nu_i}(r)$ is the luminosity as set by the energy leakage
rate interior to radius $r$, $\sigma_\mathrm{heat,\nu_i}$ is the
energy-averaged absorption cross section, $m_u$ is the atomic mass
unit, $X_i$ is the mass fraction of the absorbing particle (proton or
neutron), and $\langle 1/F_{\nu_i}\rangle$ is the mean inverse flux
factor which we approximate analytically as a function of the optical
depth $\tau$ by comparing to angle-dependent radiation transport
calculations~\cite{ott:08,oconnor:10}. The factor $e^{-2\tau_{\nu_i}}$ is used to
attenuate heating at high optical depths where neutrinos and matter
are in equilibrium. $f_\mathrm{heat}$ is a scaling factor that may be
used to dial in higher heating rates, but is set to $1$ here, causing
explosions to fail in all models. Once the heating rate is computed,
the luminosity is reduced accordingly to maintain energy conservation.

\section{NO DIRECT Black Hole Formation in Core Collapse}

The idea prevails that stars in the upper half of the $\sim 8\,
M_\odot$ to $\sim 100\,M_\odot$ mass range collapse to BHs
directly. This is incorrect, but, by an unfortunate choice of words in
recent work \cite{heger:03,whw:02} that can easily be misunderstood,
has become widely believed.

According to Thorne's Hoop Conjecture~\cite{thorne:72}, a BH forms
when a given amount of mass-energy collapses through its own
Schwarzschild radius $R_\mathrm{S} = 2 M$ (in geometric units), hence
must achieve a compactness $M/R > 0.5$. If prompt BH formation were to
occur in core collapse, the collapsing core (or parts of it) must
reach this level of compactness prior to or at bounce. In the
following, we demonstrate that this is never the case for stars in the
above mass range and explain why.

\begin{figure}[t]
\includegraphics[width=0.7\textwidth]{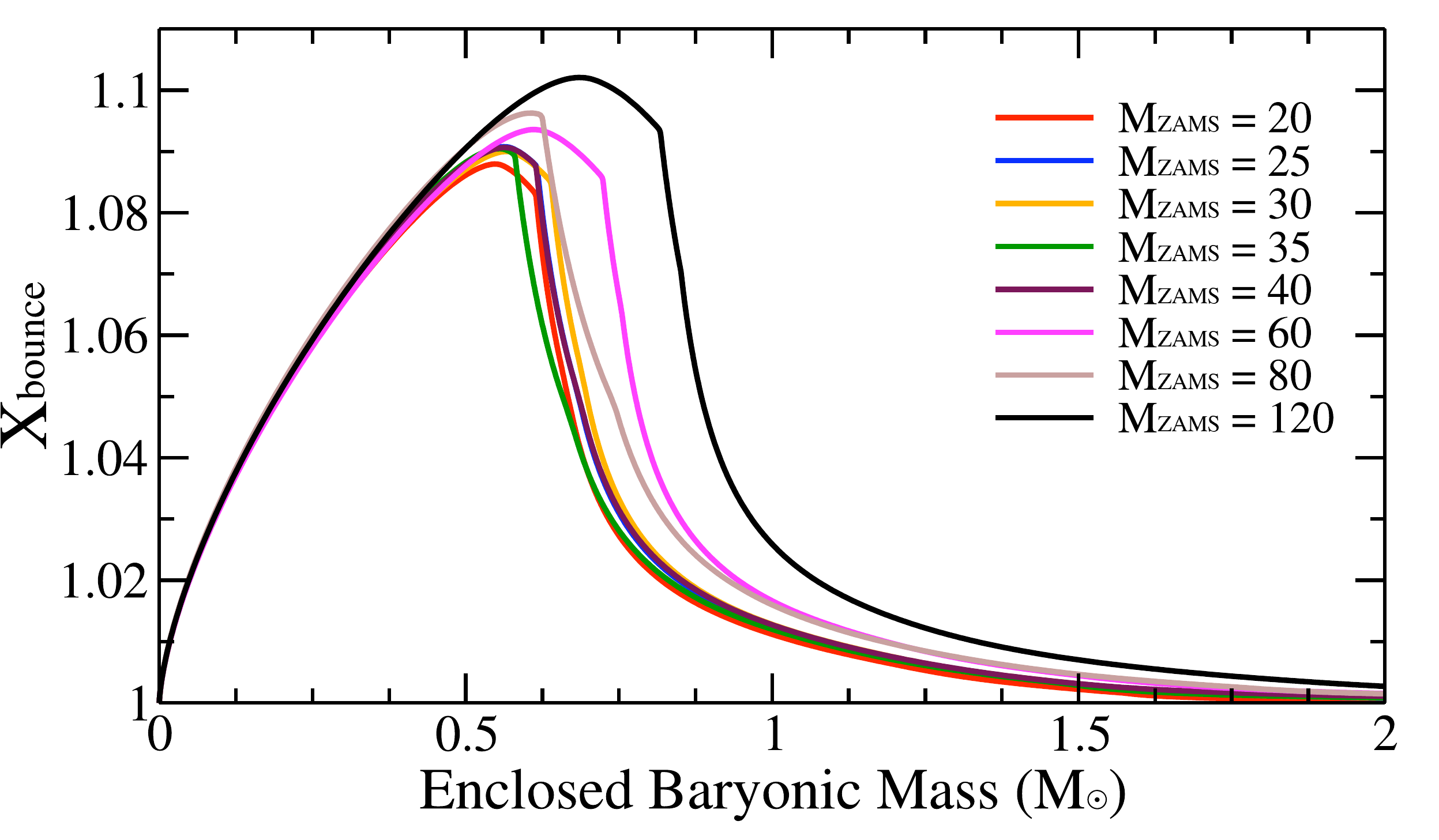}
\caption{Metric coefficient $X = 1/\sqrt{1-2m/r}$ as a function of
  enclosed baryonic mass at the time of core bounce
  and greatest inner-core compactness. 
  We plot $X$ for GR1D collapse simulations
  of models from \cite{limongi:06} with ZAMS masses from $20$ to
  $120\,M_\odot$. If a BH were to form, $X$ would diverge.
  Due to the universality of core collapse, inner core
  masses vary only little with progenitor and the compactness
  at bounce stays moderate independent of progenitor
  mass / model. There is no prompt formation of BHs.
  While GR1D's treatment of deleptonization during collapse is
  approximate, it agrees well qualitatively and quantitatively with 
  more quantitatively accurate simulations (e.g., \cite{liebendoerfer:05fakenu}).
}
\label{fig:xofm}
\end{figure}

Goldreich \& Weber~\cite{goldreich:80} and Yahil~\cite{yahil:83}
demonstrated analytically what has been confirmed numerically numerous
times: In collapse, the iron core separates into the homologously ($v
\propto r$, in sonic contact) infalling inner core (IC) and the
supersonically collapsing outer core.  It is only the inner core that
plunges to nuclear density and significant compactness in the final
phase of core collapse and it is essentially its mass
($M_\mathrm{IC}$) that must be pushed below its Schwarzschild radius
to make a BH.

$M_\mathrm{IC}$ at bounce is proportional to $Y_{e,\mathrm{IC}}^2
(1+\eta s^2_\mathrm{IC})$ \cite{burrows:83}, where $\eta \approx 0.1$
and where $Y_{e,\mathrm{IC}}$ and $s_{IC}$ are the mean $Y_e$ and
  specific entropy of the inner core at bounce, respectively.  Both
  quantities are coupled through electron capture, neutrino transport,
  and the equalibration of neutrinos and matter above trapping density.
  The physics governing $Y_{e,IC}$ and $s_\mathrm{IC}$ during collapse
  is general and independent of the conditions prevailing in a
  particular collapsing
  star~\cite{liebendoerfer:05fakenu,janka:07}. Hence, core collapse is
  universal, rather independent of initial conditions, and
  $M_\mathrm{IC}$ falls into the range of $\sim 0.4 - 0.6\,M_\odot$
  in the nonrotating case\footnote{Rotation can increase the size of
    the homologous region and, hence, $M_\mathrm{IC} $ significantly,
    but also limits inner core compactness
    \cite{dimmelmeier:08}.}~\cite{hix:03,dimmelmeier:08}. The Hoop
  Conjecture would require this mass to be compressed into
  $1.3-1.8\,\mathrm{km}$, which even for the softest plausible nuclear
  EOS~(e.g., the $K_0 = 180\,\mathrm{MeV}$ variant of the
  Lattimer-Swesty EOS~\cite{lseos:91,steiner:10}) does not occur before or
  at bounce.

In Fig.~\ref{fig:xofm}, we plot the metric coefficient $X$
(Eq.~\ref{eq:metriccoefficients}) at the time of core bounce and as a
function of enclosed baryonic mass. $X$ is shown for a set of
solar-metallicity models in the $20-120\,M_\odot$ mass range. These
are drawn from the Limongi \& Chieffi~\cite{limongi:06} model set
(using the mass-loss rates of \cite{nugis:00}) and collapsed with GR1D
without rotation, using the Lattimer-Swesty $K_0 = 220\,\mathrm{MeV}$
(LS220) EOS, and an analytic fit to the deleptonization trajectory
$Y_e(\rho)$ obtained by~\cite{liebendoerfer:05fakenu}.
\emph{The figure clearly underpins the universality of core collapse
and demonstrates that none of the collapse models come even close to BH
formation before or at bounce} (for which $X\gg1$).
Variations of the $X(m)$ profiles with progenitor mass are due
primarily to larger inner core masses owing to higher
temperatures/entropies in more massive progenitors. In a full
radiation-hydrodynamics treatment, one would expect these variations
to be even smaller due to increased deleptonization in
hotter cores.

\begin{figure}[t]
\includegraphics[width=0.7\textwidth]{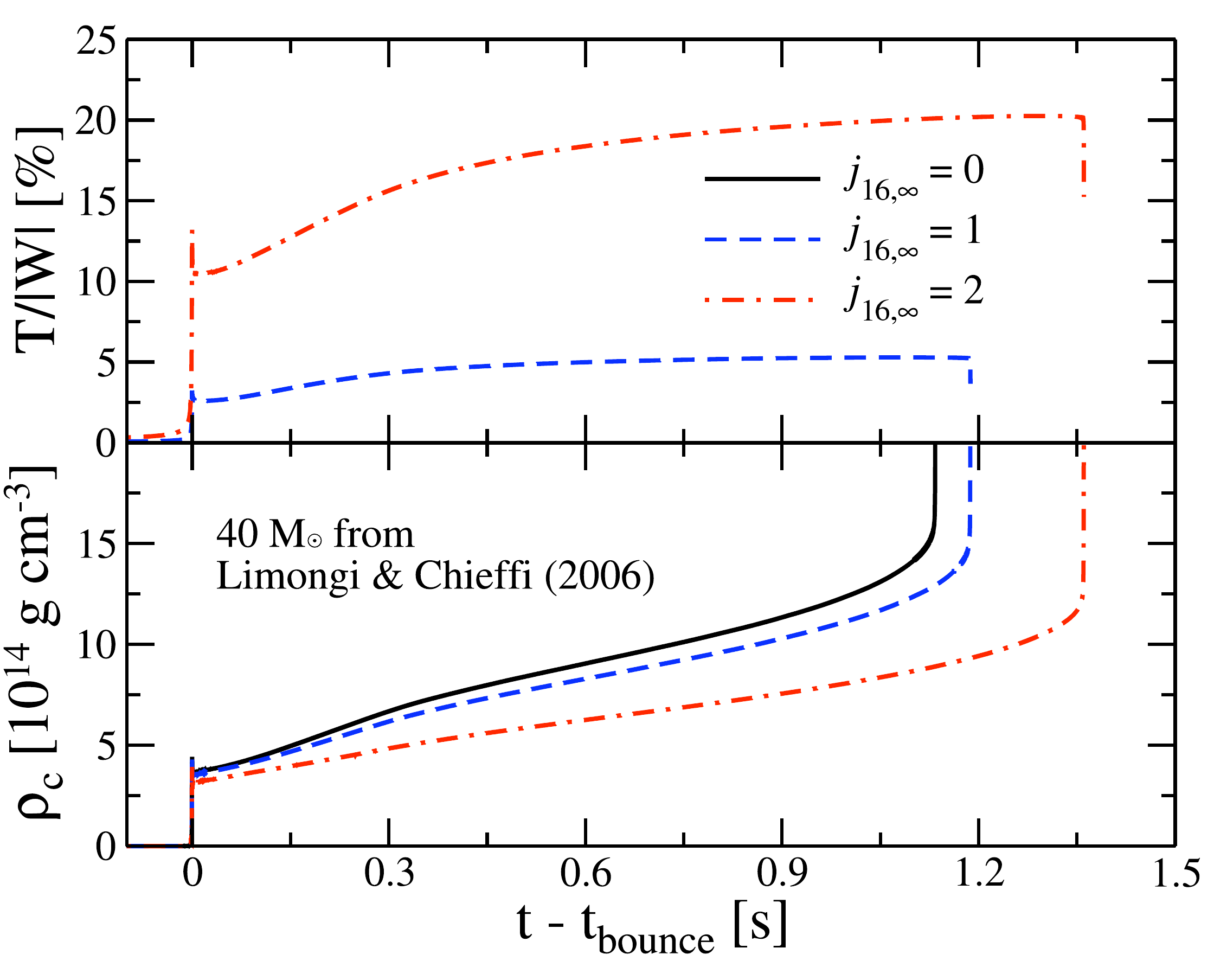}
\caption{Example results from GR1D simulations of a $40$-$M_\odot$
  progenitor of \cite{limongi:06} set up to spin with asymptotic
  specific angular momenta $j_{16} = \{0,1,2\}$ (see text for
  details). {\bf Bottom:} Evolution of the central density as a
  function of postbounce time. Rotational support leads to a lower
  accretion rate, slower PNS contraction, and a larger maximum PNS
  mass. At BH formation, the PNS baryonic mass is $2.37$, $2.39$, and
  $2.46\,M_\odot$, for the $j_{16} = 0, 1,$ and $2$ cases,
  respectively.  {\bf Top:} Ratio of rotational kinetic energy $T$ to
  gravitational binding energy $|W|$ in the two spinning
  simulations. In both models, a local maximum is reached at bounce
  and is followed by a secular increase in the postbounce phase as the
  PNS core contracts and spins up. The $j_{16} = 2$ run reaches values
  of $T/|W|$ that may make it susceptible to a secular nonaxisymmetric
  instability~\cite{lai:95} before BH formation. Both models may
  develop a low-$T/|W|$ corotation-type nonaxisymmetric instability
  (e.g., \cite{watts:05}).}
\label{fig:bhrot}
\end{figure}

\section{The Formation of Spinning Black Holes}

We perform 1.5D rotating collapse simulations of the $40$-$M_\odot$
progenitor of \cite{limongi:06} with the LS220 EOS. Rotation is added
when the peak collapse velocity of the core reaches
$1000\,\mathrm{km}\,\mathrm{s}^{-1}$ via the rotation law $\Omega(r) =
\Omega_0 ( 1 + r^2 / A^2)^{-1}$. $A$ is set to the radius at which the
enclosed baryonic mass is $1\, M_\odot$, resulting in nearly uniform
rotation throughout the inner core \cite{ott:06spin}. We perform three
calculations with $\Omega_0 = \{0,0.9,1.8\}\, \mathrm{rad\,s}^{-1}$,
corresponding to specific angular momenta at infinity in units of
$10^{16} \, \mathrm{cm}^2 \,\mathrm{s}^{-1}$ of $j_{16,\infty} =
\{0,1,2\}$.

In the lower panel of Fig.~\ref{fig:bhrot}, we show the time
evolution of the central density in the three calculations as a
function of postbounce time.  The nonrotating model forms a BH at
$\sim 1.14\,\mathrm{s}$ and has a maximum baryonic (gravitational) PNS
mass of $\sim 2.37\,M_\odot$ ($\sim 2.19\,M_\odot$). In the rotating
models, centrifugal support reduces the accretion rate and slows down
the contraction of the PNS. This alters the PNS structure and leads to
an onset of PNS collapse at later times (scaling roughly with
$\Omega_0^2$) and at lower central densities. The times of BH
formation are $\sim 1.19\,\mathrm{s}$ and $\sim 1.36\,\mathrm{s}$ in
the $j_{16,\infty} = 1$ and $j_{16,\infty} = 2$ model, respectively.
Interestingly, rotational support has only a small effect on the
maximum PNS masses in the models presented here. This is due primarily
to the fact that the PNS cores in our models are uniformly spinning. A
significant increase of the maximum PNS mass is expected only in
differentially rotating PNSs~\cite{baumgarte:00}.  We find baryonic
(gravitational) PNS masses at BH formation of $\sim 2.39\,M_\odot$
($\sim 2.22\,M_\odot$) and $\sim 2.46\,M_\odot$ ($\sim 2.30\,M_\odot$)
in the two spinning models.

The top panel of Fig.~\ref{fig:bhrot} depicts the time evolution of
the ratio of rotational kinetic energy $T$ to gravitational energy
$|W|$ in the spinning models. In both, $T/|W|$ increases secularly
after bounce as the PNS accretes and contracts and reaches a maximum
value before the onset of PNS collapse of $\sim 0.20$ ($\sim 0.05$) in
the $j_{16,\infty} = 2$ ($j_{16,\infty} = 1$) model. Both models stay
below the threshold $T/|W|_\mathrm{dyn} \approx 0.27$ for the
high-$T/|W|$ dynamical nonaxisymmetric instability (e.g.,
\cite{stergioulas:03}). The rapidly spinning PNS of the $j_{16,\infty}
= 2$ model stays above $T/|W|_\mathrm{sec} \approx 0.14$ for $\sim
1\,\mathrm{s}$, a time likely sufficiently long for a secular
(gravitational-wave or viscosity driven) nonaxisymmetric instability
to arise in 3D (e.g., \cite{lai:95}), redistributing/radiating angular
momentum and thus effectively limiting the PNS core spin. In addition,
both models may be susceptible to shear instabilities that may
generate nonaxisymmetric structure (e.g., \cite{watts:05}) and/or
magnetic flux~(e.g., \cite{obergaulinger:09}) and redistribute angular
momentum. In our present 1.5D calculations such processes and
instabilities are absent and the BH formed in the rapidly spinning
$j_{16,\infty} = 2$ model has an initial spin parameter $a^\star =
J/M_\mathrm{grav}^2$ of $\sim 0.81$. Its more slowly spinning
$j_{16,\infty} = 1$ counterpart forms a BH with $a^\star \sim 0.41$.

\section{Discussion}

In this contribution to the proceedings of the 10th International
Symposium on the Origin of Matter and Evolution of the Galaxies
(OMEG10), we have highlighted results of a small set of stellar
collapse and BH formation simulations carried out with the new
open-source 1.5D code GR1D. We numerically demonstrated what has long
been known on the basis of analytic
arguments~\cite{goldreich:80,yahil:83}, namely that a BH is never
formed promptly in the core collapse of ordinary massive stars with
masses between $\sim 8\, M_\odot$ and $\sim 100\,M_\odot$. Any core
collapse event, if ultimately resulting in BH formation or not, passes
through a protoneutron star phase in which neutrinos and, quite
likely, gravitational waves, are emitted for at least a few hundred
milliseconds. Direct subsidence into a BH is
possible only in much more massive stars that become radially unstable
before forming a hydrostatic iron core (e.g., \cite{nakazato:06}),
provided they do not experience a pair-instability driven
thermonuclear disruption instead of collapse.

Massive stars, in particular those favored as the progenitors of
long GRBs, may have rapidly spinning cores~\cite{wb:06} that will
make spinning BHs. Using GR1D's 1.5D rotation feature, we have
performed a first small set of spinning BH formation simulations.
Rotational effects increase the time to BH formation significantly
(and, thus, allow a potential explosion mechanism more time to
operate!), but, due to almost uniform rotation in the PNS core, have a
much smaller impact on the maximum PNS mass. We also find that the
rotation rate $T/|W|$ of PNSs increases significantly during the
postbounce accretion and contraction phase. PNSs with early postbounce
values of $T/|W|$ below the thresholds for secular or dynamical
rotational instabilities may surpass these within a few hundred
milliseconds. Nonaxisymmetric deformation and the associated
angular momentum redistribution and/or emission of angular momentum in
gravitational waves may put a natural limit on the maximum PNS
spin and, in consequence, on the birth spin of BHs.

\vskip.2cm
Stellar-mass BH formation is a process that occurs frequently
in the universe and core collapse is its natural site. As observations
suggest (\cite{smartt:09} and references therein), it may be the
generic ultimate outcome of core collapse in stars more massive than
$\sim 20\,M_\odot$. In the context of the core-collapse supernova --
GRB connection, BH formation is a necessary ingredient for
the collapsar scenario to work~(e.g., \cite{heger:03,wb:06}). The current
theoretical understanding of core-collapse supernovae and BH formation
is still only partial. Ultimately, it will be necessary to establish a
firm quantitative mapping between ZAMS conditions (mass, metallicity,
angular momentum) and the outcome of stellar collapse. This is a problem
as much in stellar evolution as in core-collapse supernova theory
and will require advances in both fields.

With the GR1D code we have created an open-source tool that allows us
and others to study BH formation in failing core-collapse
supernovae and to investigate its systematics and the characteristics
of nascent BHs with variations in presupernova stellar structure and
rotational configuration. While a quantitatively robust mapping to
ZAMS conditions may not be possible on the basis of currently
available stellar evolutionary models, robust qualitative features and
trends can be derived in parameter studies \cite{oconnor:10b}.

\begin{theacknowledgments}
  It is a pleasure to thank the organizers of the OMEG10 symposium.
  Furthermore, we acknowledge helpful and stimulating conversations
  with W.~D.~Arnett, A.~Burrows, M.~Duez, T.~Fischer, C.~Fryer,
  J.~Lattimer, C.~Meakin, F.~Peng, C.~Reisswig, E.~Schnetter, H.~Shen,
  U.~Sperhake, K.~Sumiyoshi, F.~Timmes, K.~Thorne, and H.~Toki.  CDO
  is supported in part by the National Science Foundation under grant
  numbers AST-0855535 and OCI-0905046.  EOC is supported in part
  through a post-graduate fellowship from the Natural Sciences and
  Engineering Research Council of Canada (NSERC). Computations were
  performed on the Louisiana Optical Network Infrastructure computer
  systems under allocation loni\_numrel04.
\end{theacknowledgments}

\end{document}